\documentclass{article}
\usepackage{spconf,amsmath,graphicx}
\usepackage{amsfonts}                               
\DeclareMathOperator*{\argmax}{arg\,max}            
\usepackage{multirow}                               
\usepackage{caption}                                
\title{Unsupervised data selection for Speech Recognition with contrastive loss ratios}
\name{Chanho Park, Rehan Ahmad, Thomas Hain}
\address{Department of Computer Science, University of Sheffield,\\ cpark12, r.ahzzmad, t.hain@sheffield.ac.uk}
\begin{document}
%
\maketitle
\begin{abstract}
This paper proposes an unsupervised data selection method by using a submodular function based on contrastive loss ratios of target and training data sets. A model using a contrastive loss function is trained on both sets. Then the ratio of frame-level losses for each model is used by a submodular function. By using the submodular function, a training set for automatic speech recognition matching the target data set is selected. Experiments show that models trained on the data sets selected by the proposed method outperform the selection method based on log-likelihoods produced by GMM-HMM models, in terms of word error rate (WER). When selecting a fixed amount, e.g. 10 hours of data, the difference between the results of two methods on Tedtalks was 20.23\% WER relative. The method can also be used to select data with the aim of minimising negative transfer, while maintaining or improving on performance of models trained on the whole training set. Results show that the WER on the WSJCAM0 data set was reduced by 6.26\% relative when selecting 85\% from the whole data set. 
\end{abstract}
\section{Introduction}
\label{sec:intro}
The performance of an automatic speech recognition (ASR) system is affected by the amount of training data used. It would be ideal to use more data with transcriptions. Unfortunately, since labelling is time-consuming, unlabelled data are used as supplementary to the labelled data set \cite{parthasarathi2019lessons}. The above method using both labelled and unlabelled data sets is called semi-supervised learning. Furthermore, an unsupervised speech recognition method that does not require labelled data \cite{baevski2021unsupervised} has been investigated to gain more information from unlabelled data. As a result, the training time has increased as the amount of unlabelled data required for these methods has increased. 

The semi-supervised learning method has shown improvement in WER in both cases of using additional data in a single domain \cite{zhang2020pushing} and multi-domain data consisting of various data sets \cite{hsu21_interspeech}. In \cite{hsu21_interspeech}, the model trained on a multi-domain training data set outperformed the model trained on a single-domain data set. However, in the case of multi-domain data, the degradation of WER performance between different data sets, called negative transfer, was observed in \cite{doulaty15_interspeech}. For example, the WER of the model trained on both Fisher and WSJCAM0 data sets was worse than that of the model trained on one of the data sets. Moreover, a similar observation exists in \cite{hsu21_interspeech}. More subsets in pre-training showed mixed results in terms of WER performance in the scenario when target data were included in the training set. 

So far, two observations have been discussed: increased training time and negative transfer. Given a budget of the amount of training data, they need to be addressed by selecting an optimal data subset without performance degradation. Confidence scores can be used to select only high confidence data. However, it is time-consuming because they are built upon the top of ASR systems \cite{li2021confidence}. Second, to address the computational cost of confidence scores, a proxy function which is smaller but faster has been proposed \cite{coleman2019selection}. However, the performance of a proxy model is less accurate than that of the original as it is proposed to reduce the computational cost instead of the accuracy. There has been another approach to data selection using a submodular function in a budget \cite{lin2009select, wei2014unsupervised}. A submodular function converts a set into a measurable value so that the optimal subset in the budget can be found by the value. 

In this paper, an unsupervised submodular data selection method by using contrastive loss ratios is proposed. The proposal is based on a contrastive loss function learned by predicting future latent representations conditioned on past latent representations \cite{oord2018representation, schneider19_interspeech}. After pre-training, the loss of the model trained on a target data set indicates how likely the frame is in the context of the data set. The loss is compared to that of the model trained on a training data set. Whether the frame is close to the target data or to the training data is decided by the ratios between them. By the utterance-level mean value of the frame-level ratios, all the utterances can be ordered. A function for a contrastive loss ratio is monotonic and non-decreasing by adding utterances. With the characteristics of the function, it can be used as a submodular function for selecting an optimal subset. 

The remaining part of the paper proceeds as follows: in Section \ref{sec:Background}, background information is introduced. In Section \ref{sec:Proposed method}, the proposed approach is described in detail. After the experimental setup in Section \ref{sec:Experimental setup}, the results are shown in Section \ref{sec:Results}. Then, the conclusion is in Section \ref{sec:Conclusion}. 

\section{Background}
\label{sec:Background}

\subsection{Contrastive representation learning}
\label{ssec:Contrastive representation learning}
A contrastive loss function maximises the similarity between data representations in a category while minimising it in different categories. 

A representation learning method by using the contrastive loss function has been proposed in \cite{oord2018representation,schneider19_interspeech}. First, an audio frame $x_t$ at time $t$ is embedded by convolutional neural networks. Next, an embedding $z_t$ is contextualised by using an autoregressive model to exploit high-level latent information between different parts. Then, the context representation $c_t$ is used to predict future embeddings $z_{t+k}$ where $k$ is the number of future steps. 

Instead of predicting $x_{t+k}$ directly, the mutual information of $x_{t+k}$ and $c_t$ is maximised. Based on the model, InfoNCE is suggested as an objective function in \cite{oord2018representation}. Both the encoder and autoregressive model are trained to jointly optimise the InfoNCE loss: 
\begin{equation}
\label{eq:InfoNCE}
    \mathcal{L}_N = -\mathbb{E}_X \Big[ log \frac{f_k(x_{t+k}, c_t)}{\sum_{x_j \in X}f_k(x_j, c_t)} \Big]
\end{equation}
where $f$ is a density ratio and can be unnormalised, and $X$ is a set of random samples including a positive sample. To score the mutual information, a simple log-bilinear model was used in \cite{oord2018representation}. Similarly, the wav2vec model in  \cite{schneider19_interspeech} used $\sigma(z^\top_{t+k} h_k(x_t))$ as $f$ where $\sigma$ is a sigmoid function, and $h_k$ is an affine transformation. 

\subsection{Submodular function maximisation}
\label{ssec:Submodular maximisation}

Selecting data from a data pool is to find discrete sets of feasible solutions. This problem can be solved by adopting a submodular function \cite{krause2014submodular} for the set. 
\begin{equation}
\label{eq:Submodular function}
    f: 2^V \rightarrow \mathbb{R}
\end{equation}
where $V$ is a finite set, $2^V$ is the set of all subsets of $V$ and $f(\emptyset) = 0$. $f$ is submodular if $f_A(e) \geq f_B(e)$ for all $A \subseteq B \subseteq V$ and $e \in V \backslash B$ where $f_A(e) = f(A \cap \{e\}) - f(A)$.
If the function is concave, the optimal solution $S$ is a set which maximises the value of the submodular function. 
\begin{equation}
\label{eq:Submodular maximisation}
    \text{max}\{f(S): S \subseteq V\}
\end{equation}
To solve this problem, a greedy solution can be used. However, it is NP-hard and requires the non-deterministic polynomial time. One possible idea to avoid it is for the function to be monotonic. 
\begin{equation}
\label{eq:Submodular monotonic}
    f(A) \le f(B) \text{, if } A \subseteq B \subseteq V
\end{equation}
With monotonic functions, it is clear that $f$ is maximised at V. Now the optimal solution can be found by considering a constraint $k$ for data selection. 
\begin{equation}
\label{eq:Submodular constraint}
    \argmax_{|S| \le k}\{f(S)\}
\end{equation}
In other words, each element of a set can be selected by the value of the function in order. 

\section{Proposed method}
\label{sec:Proposed method}

\subsection{Loss ratios}
\label{ssec:Loss ratios}
As a submodular function, the mean of the ratios of all the frames in an utterance is used. First, ratios between the values of a contrastive loss function $f_\Omega$ trained on the whole training data set $\Theta_{\Omega}$ and those of another contrastive loss function $f_{tgt}$ on a target data set $\Theta_{tgt}$ can be calculated as below: 

\begin{equation}
\label{eq:Loss ratios}
    LR(u) = \frac{1}{T}\sum_{t=1}^T \frac{f_\Omega(x_t) + \alpha}{f_{tgt}(x_t) + \alpha}
\end{equation}
where $\alpha$ is a number to prevent overflow or underflow of the loss ratio and $x_t$ is an observation at time $t$. Then, the accumulated $LR(u)$ of all utterances in a subset $S$ which is included in $\Omega$ is defined as a submodular function like below: 

\begin{equation}
\label{eq:Submodular function}
    f_{LR}(S) = \sum_{u \in S} \big( LR(u) \big)
\end{equation}

This modular function $f_{LR}$ is non-negative as it is a sum of means of ratios between non-negative losses in Equation \ref{eq:InfoNCE}. Moreover, it is normalised, which means that the value of the function is zero when the input is null. 

\subsection{Negative transfer minimisation}
\label{ssec:Negative transfer minimisation}

When a budget is not given, the amount of data can be reduced without degradation of WER performance by minimising the negative transfer. Negative transfer is performance degradation when another domain data set is added to a training set. To maximise the performance of an ASR model, negative transfer should be minimised. As the optimal data subset is selected by the value of the submodular function in Equation \ref{eq:Submodular function}, a data set of negative transfer can be filtered by a threshold. The threshold is investigated by a grid search method. Furthermore, a value of $f_{tgt}(x_t)$ represents how well the model is fitted into the target data set. If an utterance is affected by negative transfer, $f_{tgt}(x_t)$ would be high. Thus, another threshold of loss is also explored in the same method. 

\section{Experimental setup}
\label{sec:Experimental setup}

Similar to the previous study \cite{doulaty15_interspeech}, a data pool (DP) was generated to select from. It consisted of 4 data sets: AMI \cite{cieri2004fisher}, Fisher \cite{carletta2005ami} (FS), Tedtalks \cite{ng2015usfd} (TD) and Wsjcam0 \cite{robinson1995wsjcamo} (WS0). The training data pool was 40 hours in total: 10 hours of each data set. For pre-training, two 1-hour sets of each data set were target and test data sets. There are three parts of the experiment: pre-training; data selection; building ASR systems. 

\subsection{Pre-training}
\label{ssec:Pre-training}

A wav2vec \cite{schneider19_interspeech} was modified for contrastive loss ratios. To fit into the small size of the target data set, kernel sizes and strides of encoder layers were changed from (10,8,4,4,4) and (5,4,2,2,2) to (16, 16) and (2,3,4), respectively. The number of prediction steps was set to 6 instead of 12. Each model was trained to 200 epochs, then stopped early with patience of 10 epochs. Each model was trained on individual target and training data sets.


A GMM-HMM system was trained for log-likelihood. As the amount of training data for this purpose was an hour, the accuracy of the system was not as high as the hybrid ASR system described in Section \ref{ssec:Hybrid ASR system}. Then, the training data were decoded by using the models, and the log-likelihood of the utterances in the decoding was applied for data selection. 
 
\subsection{Data selection}
\label{sssec:Data selection}

After pre-training, frame-level losses on a target data set and the training data set were calculated. The losses were added to $\alpha$ to prevent ratios of them from overflowing. Then, the mean values of the losses of an utterance were used as a score for data selection. The data were selected by a greedy method based on the score. The utterances were sorted by the score, then data in a budget was selected in order. When a budget was not given, contrastive loss ratios and losses were used to reduce the amount of the selected data by minimising negative transfer. The thresholds of contrastive loss ratios and losses were investigated empirically. 

\subsection{Hybrid ASR system}
\label{ssec:Hybrid ASR system}

Hybrid systems of GMM-HMM and neural networks were built for speech recognition. They were based on the system described in \cite{povey2014parallel}. For alignment, monophone, triphone, discriminant analysis (LDA) and maximum likelihood linear transform (MLLT), and Speaker Adaptive Training (SAT) models were trained in sequence. Then, neural networks (Nnet2) were trained by using the labelled frames generated by the GMM-HMM models. Moreover, the language model was trained on merged text of all data sets. 
\section{Results}
\label{sec:Results}

\subsection{Baseline}
\label{ssec:Baseline}

An ASR system with the data pool was built as a baseline system. It was trained on all 40 hours of the data pool, and each test data set and the combined data set (DP) were scored separately as seen in Table \ref{table:WERs of baseline systems}. 

\begin{table}[htbp]
\caption{WERs(\%) of baseline systems}
\label{table:WERs of baseline systems}
    \begin{center}
    \begin{tabular}{ c|c|c|c|c|c }
     \hline
     Feature & AMI & FS & TD & WS0 & DP \\ 
     \hline
     \hline
     MFCC & 26.69 & 35.72 & 24.58 & 9.90 & 25.04 \\ 
     \hline
    \end{tabular}
    \end{center}
\end{table}

\subsection{Data selection}
\label{ssec:Data selection}

The results of data selection are shown in Table \ref{table:Selected data by contrastive loss ratios}. The first column is a target data set which a pre-training model was trained on. When segments were selected by the contrastive loss ratios between a pre-training model on AMI and another model on the data pool, 3263, 3503 and 3521 segments of AMI were selected for sets of 10h, 20h and 30h from the pool, respectively. At the same time, 2023, 2810 and 3222 segments were selected by log-likelihood, respectively. Data from the same corpus with the target data tended to be selected by contrastive loss ratios rather than by log-likelihood. 

\begin{table}[htbp]
\caption{Numbers of selected segments by contrastive loss ratios and log-likelihood. The total numbers for AMI, FS, TD and WS0 were 3526, 3330, 3244 and 3685, respectively. CLR and LL stand for contrastive loss ratio and log-likelihood, respectively.}
\label{table:Selected data by contrastive loss ratios}
    \begin{center}
    \begin{tabular}{ c|c|c|c|c }
     \hline
     \multirow{2}{3.5em}{target\\data set} & \multicolumn{3}{|c|}{hours of subset (CLR/LL)} & \multirow{2}{3.5em}{selected\\data set}\\ 
     \cline{2-4}
     & 10h & 20h & 30h &  \\ 
     \hline
     \hline
     \multirow{4}{3em}{AMI} & 3263/2023 & 3503/2810 & 3521/3222 & AMI \\
     & 14/131 & 291/774 & 1083/1863 & FS \\
     & 195/306 & 1811/1089 & 2725/2020 & TD \\
     & 16/1008 & 1320/2261 & 3070/3262 & WS0 \\
     \hline
     \multirow{4}{3em}{FS} & 0/13 & 669/1616 & 2209/2717 & AMI \\  
     & 3257/3301 & 3328/3325 & 3329/3325 & FS \\
     & 65/18 & 2615/1399 & 3123/2455 & TD \\
     & 0/0 & 15/349 & 1479/1646 & WS0 \\
     \hline
     \multirow{4}{3em}{TD} & 103/1385 & 1524/2250 & 2797/2899 & AMI \\  
     & 362/162 & 1789/781 & 2686/1807 & FS \\
     & 2773/1100 & 3181/2099 & 3219/2779 & TD \\
     & 0/720 & 152/1662 & 1471/2781 & WS0 \\
     \hline
     \multirow{4}{3em}{WS0} & 104/845 & 2166/2492 & 3299/3208 & AMI \\  
     & 0/4 & 4/337 & 334/1699 & FS \\
     & 28/57 & 1222/625 & 3116/1861 & TD \\
     & 3527/2680 & 3684/3653 & 3685/3685 & WS0 \\
     \hline
    \end{tabular}
    \end{center}
\end{table}

Given 10, 20, 30 hours budgets, most of the selected data for the first 10 hours were from the same data set as the target data. As the budget increases, data from the different data sets were selected more because the amount of each data set was limited to 10 hours. However, one of the data sets was less frequently selected until 30 hours. For instance, when a target data set was WS0, only 334 utterances of FS were selected for the 30-hour data set. This was relatively lower than the other data selection results. The negative effect of the remaining data was observed in the following section by using the WER performance. 

\subsection{Loss ratios vs log-likelihood}
\label{ssec:Loss ratios vs likelihood}
Data selected sets in Table \ref{table:Selected data by contrastive loss ratios} were used for training ASR models. The performance of each system was measured in terms of WER. The results of ASR systems trained on data selected by contrastive loss ratios outperformed those by log-likelihood of the hybrid system. For example, in Table \ref{table:WERs on selected data sets}, the WER performances of models trained on data sets by contrastive loss ratios outperformed the others except 20 hours and 30 hours when the target sets were FS and AMI, respectively. 

\begin{table}[htbp]
\caption{WERs(\%) on selected data sets. For example, the WER of an ASR system trained on the subset of 10h when a target data set was AMI was 31.71\%.}
\label{table:WERs on selected data sets}
    \begin{center}
    \begin{tabular}{ c|c|c|c|c|c }
     \hline
     Method & target & 10h & 20h & 30h & 40h\\ 
     \hline
     \hline
     \multirow{4}{3em}{CLR} & AMI & 31.71 & 28.62 & 27.02 & \textbf{26.69} \\
     & FS & 39.57 & 37.12 & \textbf{35.49} & 35.72 \\
     & TD & 28.07 & 25.54 & \textbf{24.43} & 24.58 \\
     & WS0 & 11.14 & 9.57 & \textbf{9.32} & 9.90 \\
     \hline
     \multirow{4}{3em}{LL} & AMI & 34.51 & 29.56 & 26.95 & \textbf{26.69} \\
     & FS & 40.02 & 36.80 & 36.56 & \textbf{35.72} \\
     & TD & 35.19 & 28.37 & 26.42 & \textbf{24.58} \\
     & WS0 & 11.27 & 9.90 & \textbf{9.89} & 9.90 \\
     \hline
    \end{tabular}
    \end{center}
\end{table}

\subsection{Negative transfer minimisation}
\label{ssec:Negative transfer minimisation}

For a grid search of thresholds, different amounts of data sets were selected by contrastive loss ratios and losses. Based on the result in Table \ref{table:WERs on selected data sets}, from 80\% to 95\% of the data pool were selected as training data sets. 

\begin{table}[htbp]
\caption{WERs(\%) on selected data sets for negative transfer minimisation by contrastive loss ratios and losses. CL stands for contrastive loss.}
\label{table:negative transfer minimisation}
    \begin{center}
    \begin{tabular}{ c|c|c|c|c|c }
     \hline
     Method & target data set & 80\% & 85\% & 90\% & 95\% \\
     \hline
     \hline
     \multirow{4}{3em}{CLR} & AMI & 26.98 & 26.79  & \textbf{25.91} & 26.35 \\
     & FS & 35.83 & 36.96 & 35.83 & \textbf{35.72} \\
     & TD & 24.97 & 25.25 & 24.94 & \textbf{24.34} \\
     & WS0 & 9.66 & 9.71 & \textbf{9.51} & 9.66 \\
     \hline
     \multirow{4}{3em}{CL} & AMI & 27.19 & 26.55 & \textbf{25.78} & 27.36 \\
     & FS & \textbf{35.02} & 36.11 & 35.75 & 35.50 \\
     & TD & 25.09 & 24.61 & \textbf{24.34} & 24.59 \\
     & WS0 & 9.56 & \textbf{9.28} & 9.66 & 9.52 \\
     \hline
    \end{tabular}
    \end{center}
\end{table}

As shown in Table \ref{table:negative transfer minimisation}, the optimal subsets for WER performance were between 80\% and 95\% of the whole data set. The best performances for AMI, FS, TD and WS0 on data sets selected by CLR were 25.91\%, 35.72\%, 24.34\% and 9.51\%, respectively. When data sets were selected by CL, they were 25.78\%, 35.02\%, 24.34\% and 9.28\% for AMI, FS, TD and WS0, respectively. These WERs were competitive to the baseline with the CLR method when budgets were given in Table \ref{table:WERs on selected data sets}. For example, when data sets selected by CL, the WER of a model on the 80\% data set for FS were 35.02\%, while the baseline performance was 35.72\%. In other words, the WER performance on the data pool can be achieved on a smaller data set by eliminating data that might cause negative transfer. 

\section{Conclusion}
\label{sec:Conclusion}

An unsupervised data selection with a submodular function based on contrastive loss ratios has been explored in this paper. For data selection based on budget, this method outperforms a method using log-likelihood produced by the models trained on a target data set. Furthermore, the amount of data selected from the data pool can be minimised without performance degradation by avoiding negative transfer. The performance of the ASR models trained on the training data set selected by contrastive losses and ratios is comparable to the baseline performance on the whole training data set. 

\section{Acknowledgements}
\label{sec:Acknowledgements}

This work was conducted at the VoiceBase Research Centre for Speech and Language Technologies at the University of Sheffield, which is funded by VoiceBase Inc. 

\vfill\pagebreak

\bibliographystyle{IEEEbib}
\bibliography{citations}

\end{document}